\documentclass{article}
\usepackage{epsfig}
\begin{document}

\title{{\bf HLA and HIV Infection Progression: Application of the Minimum Description Length Principle to Statistical Genetics}}

\author{Peter T. Hraber$^{\ast,\dag}$, 
Bette T. Korber$^{\ast,\dag}$, 
Steven Wolinsky$^\ddag$,\\
Henry Erlich$^\S$, 
Elizabeth Trachtenberg$^\P$, and 
Thomas B. Kepler$^{\ast,||}$\\
\\
\normalsize $^\ast$Santa Fe Institute, 1399 Hyde Park Road, Santa Fe NM 87501\\
\normalsize $^\dag$Los Alamos National Laboratory, Los Alamos NM 87545\\
\normalsize $^\ddag$Feinberg School of Medicine, Northwestern University,\\
\normalsize 676 North St. Claire, Suite 200, Chicago IL 60611\\
\normalsize $^\S$Roche Molecular Systems, 1145 Atlantic Avenue, Alameda CA 94501\\
\normalsize $^\P$Children's Hospital Oakland Research Institute,\\
\normalsize 5700 Martin Luther King Jr. Way, Oakland CA 94609\\
\normalsize $^{||}$Department of Biostatistics and Bioinformatics, and\\
\normalsize Center for Bioinformatics \& Computational Biology,\\
\normalsize Box 90090, Duke University, Durham NC 27708\\
}
\date{}
\maketitle

\thispagestyle{empty}
\normalsize

\begin{center}
\begin{description}

\item {\bf Classification}\\
Biological Science/Immunology \& Physical Science/Applied Mathematics\\

\item {\bf Corresponding author}\\
Peter T. Hraber\\
address: Santa Fe Institute, 1399 Hyde Park Road, Santa Fe NM 87501\\
phone: +1 (505) 984-8800\\
fax: +1 (505) 982-0565\\
email: pth@santafe.edu\\

\item {\bf Manuscript information}\\
Text pages: 14\\
Figures: 1\\
Tables: 2\\
Words in abstract: 245 ($<250$)\\
Character count: $45536$ ($<47000$)\\

\item {\bf Nonstandard abbreviations}\\
MACS: multicenter AIDS cohort study\\
MDL: minimum description length\\

\end{description}
\end{center}
\newpage

\vspace{3 in}
\begin{abstract}

The minimum description length (MDL) principle was developed in the
context of computational complexity and coding theory. It states that
the best model to account for some data minimizes the sum of the
lengths, in bits, of the descriptions of the model and the data as
encoded via the model. The MDL principle gives a criterion for
parameter selection, by using the description length as a test
statistic. Class I HLA genes play a major role in the immune response
to HIV, and are known to be associated with rates of progression to
AIDS. However, these genes are highly polymorphic, making it difficult
to associate alleles with disease outcome, given statistical issues of
multiple testing. Application of the MDL principle to immunogenetic
data from a longitudinal cohort study (Chicago MACS) enables
classification of alleles associated with plasma HIV RNA abundance, an
indicator of infection progression. We recently reported that MDL
analysis of the relationship of HLA supertypes (a classification of
alleles by epitope-binding anchor motifs) with HIV RNA levels
identifies associations between human genotype and viral RNA. Details
of the MDL approach and more extended analyses of HLA and viral RNA
are described here. Variation in progression is strongly associated
with HLA-B. Allele associations with viral levels support and extend
previous studies. In particular, individuals without {\em B58s}
supertype alleles average viral RNA levels 3.6-fold greater than
individuals with them. Mechanisms for these associations include
variation in epitope specificity and selection that favors rare
alleles.
\end{abstract}
\newpage

Progression of HIV infection is characterized by three phases: acute,
or early, chronic, and AIDS, the final phase of infection preceeding
death \cite{McMichael01}. The chronic phase is variable in duration,
lasting ten years on average, but varying from two to twenty years.  A
good predictor of the duration of the chronic phase is the viral RNA
level during chronic infection, with higher levels consistently
associated with more rapid progression than lower levels
\cite{Mellors96}. A major challenge for treating HIV and developing
effective vaccination strategies is to understand what contributes to
variation in plasma viral RNA levels, and hence to infection
progression.

The cell-mediated immune response identifies and eliminates infected
cells from an individual. A central role in this response is played by
the major histocompatibility complex (MHC), in humans, also known as
human leukocyte antigens (HLA). Two classes of HLA genes code for
co-dominately expressed cell-surface glycoproteins, and present
processed peptide to circulating T-cells, which discriminate between
self and non-self \cite{Germain,WilliamsReview}.

Class I HLA molecules are expressed on all nucleated cells except germ
cells. In infected cells, they bind and present antigenic peptide
fragments to T-cell receptors on CD8$^+$ T-lymphocytes, which are
usually cytotoxic and cause lysis of the infected cell. Class II HLA
molecules are expressed on immunogenetically reactive cells, such as
dendritic cells, B-cells, macrophages, and activated T-cells. They
present antigen peptide fragments to T-cell receptors on CD4$^+$
T-lymphocytes and the interaction results in release of cytokines that
stimulate the immune response.

Human HLA loci are among the most diverse known \cite{Bodmer,Little}.
This diversity provides a repertoire to recognize evolving antigens
\cite{Little,Hill}. Previous studies of associations between
HLA alleles and variation in progression of HIV-1 infection have
established that within-host HLA diversity helps to inhibit viral
infection, by associating degrees of heterozygosity with rates of HIV
disease progression \cite{Roger}. Thus, homozygous individuals,
particularly at the HLA-B locus, suffer a greater rate of progression
than do heterozygotes \cite{Roger,Carrington99}.  Identifying which
alleles are associated with variation in rates of infection
progression has been difficult, due in part to the compounding of
error rates incurred when testing many alternative hypotheses, and
published results do not always agree \cite{otherMS,Trachtenberg01}.

This study demonstrates the use of an information-based criterion for
statistical inference. Its approach to multiple testing differs from
that of standard analytic techniques, and provides the ability to
resolve associations between variation in HIV RNA abundance and
variation in HLA alleles.

As an application of computational complexity and optimal coding
theory to statistical inference, the minimum description length (MDL)
principle states that the best statistical model, or hypothesis, to
account for some observed data is the model that minimizes the sum of
the number of bits required to describe both the model and the data
encoded via the model \cite{Rissanen,Li93,HansenYu}. It is a
model-selection criterion that balances the need for parsimony and
fidelity, by penalizing equally for the information required to
specify the model and the information required to encode the residual
error.

The analyses detailed below apply the MDL principle to the problem of
partitioning individuals into groups having similar HIV RNA levels,
based on HLA alleles present in each case.

\subsection*{Chicago MACS HLA \& HIV Data}

The Chicago Multicenter AIDS Cohort Study (MACS) provided an
opportunity to analyze a detailed, long-term, longitudinal set of
clinical HIV/HLA data \cite{otherMS}. Each participant provided
informed consent in writing. Of 564 HIV-positive cases sampled in the
Chicago MACS, 479 provided information about both the rate of disease
progression and HLA genetic background. Progression was indicated by
the quasi-stationary ``set-point'' viral RNA level during chronic
infection. Immunogenetic background was obtained by determining which
HLA alleles from class I (HLA-A, -B, and -C) and class II (HLA-DRB1,
-DQB1, and -DPB1) loci were present in each individual.

Viral RNA set-point levels were determined after acute infection and
prior to any therapeutic intervention or the onset of AIDS, as defined
by the presence of an opportunistic infection or CD4$^+$ T-cell count
below 200 per ml of plasma. Because the assay has a detection
threshold of 300 copies of virus per ml \cite{otherMS},
maximum-likelihood estimators were adjusted to avoid biased estimates
of population parameters from a truncated, or censored, sample
distribution \cite{normal}. Viral RNA levels were log-transformed so
as better to approximate a normal distribution.

High-resolution class I and II HLA genotyping \cite{otherMS} provided
four-digit allele designations, though analyses were generally
performed using two-digit allele designations because of the resulting
reduction of allelic diversity and increased number of samples per
allele. Because of the potential for results to be confounded by an
effect associated with an individual's ethnicity or revised sampling
protocol, two separate analyses were performed, one using data from
the entire cohort, and another using only data from Caucasian
individuals. Sample numbers were too small to study other subgroups
independently.

HLA supertypes group class I alleles by their peptide-binding anchor
motifs \cite{supertypes}. Assignment of four-digit allele designations
to functionally related groups of supertypes at HLA-A and -B loci
facilitated further analysis. Where they could be determined, HLA-A
and HLA-B supertypes were assigned from four-digit allele designations
\cite{otherMS}. As with two-digit allele designations for each locus,
HLA-A and -B supertypes were assessed for association with viral RNA
levels. Cases having other alleles were withheld from classification
and subsequent analysis of supertypes.

A description length analysis determined whether HIV RNA levels were
non-trivially associated with alleles at any HLA locus.

\subsection*{Description Lengths}

The challenge of data classification is to find the best partition,
such that observations within a group are well-described as
independent draws from a single population, but differences in
population distributions exist between groups. Whether the data are
better represented as two groups, or more, than as one depends on the
description lengths that result.

We use the family of Gaussian distributions to model viral RNA
levels. While the MDL strategy can be applied using any probabilistic
model, a log-normal distribution is a good choice for the observed
plasma viral RNA values. First, the description length of the model
and of the data given the model is calculated as described below,
grouping all of the observations into one normal distribution,
$L_1$. Next, the data are broken into two partitions, $L_2$, and
the log-RNA values associated with HLA alleles are partitioned to
minimize the description length given the constraint that two Gaussian
distributions, each having their own mean and variance, are used to
model the data.

For fixed $n \times n$ covariance matrix $\Sigma$, the description
length is $L_\Sigma = \frac{1}{2}\log |\Sigma| + \frac{1}{2}
Y'\Sigma^{-1}Y + C$, where $Y$ is the $n$-component vector of
observations and $C$ is the quantity of information required to
specify the partition. Logarithms are computed in base two, with
fractional values rounded upwards, so that the resulting units are
bits. The description length of interest results from integrating $L$
over all covariance matrices with the appropriate structure. In
practice, we use Laplace's approximation for the integral
\cite{Rissanen,Lindley} which gives, asymptotically, $L =
\frac{1}{2}\log |\hat{\Sigma}| + \frac{1}{2} Y'\hat{\Sigma}^{-1}Y +
\frac{k}{2} \log n + C$, where $k$ is the number of free parameters in
the covariance model, and $\hat{\Sigma}$ is the specific covariance
matrix of the appropriate structure that minimizes $L_\Sigma$.
A more detailed account appears in the Appendix.

The analog of a null hypothesis is the assumption that one group of
alleles is sufficient to account for the variation in viral RNA. The
description length for one group is: $L_1=\frac{1}{2}\left(n+(n-1)\log
s^2 +\log n\overline{x}^2+2\log n\right)$, where $n$ is the total
number of observations, $s^2$ is the maximum-likelihood estimate of
the population variance and $\overline{x}$ is the sample mean,
computed as the Winsorized mean \cite{normal} because of truncation
below the sensitivity limit of the RNA assay.



It follows that the description length for two groups can be computed
as:

\[ L_{2}=\frac{1}{2} \sum_{i=1}^2 \left( n_i + (n_i-1)\log s_i^2 + \log n_i\overline{x}_i^2 + 2 \log n_i \right) + C,\]

where $C$ is an adjustment for performing multiple comparisons.
Because additional information is required to specify the optimum
partition, the description length is increased by a quantity related
to the number of partitions evaluated, such that $C = N \log k$ bits,
where $N$ is the number of alleles observed at the partitioned
locus. For $k=2$, $C=N$.

Further partitions of alleles into more than two groups might yield a
shorter description length, computed as a summation over terms in the
equation for $L_{2}$ for each of the $k$ distinct groups.

The shortest description length for any value of $k$ indicates the
best choice of model parameters, including the number of parameters,
and hence, the optimum partition of $N$ alleles into $k$ groups. We
denote this as $L^*$.

\subsection*{Algorithm}

The minimum description length is found by iteratively computing the
description length for each possible partition of alleles into groups
and taking the minimum as optimal. Iteration consists first of
determining the number of alleles, $N$, at a particular locus, and
then incrementing through each of the $k^{(N-1)}$ possible partitions
of alleles into $k$ groups, computing the associated description
length, and reporting the best results. Each iteration evaluates one
possible mapping of alleles to groups. Searching through all possible
partitions using the description length as an optimality criterion
ensures selection of the best partition as a result of the search.

In this mapping, the ordering of groups is informative, because the
ordering gives the relative dominance of alleles for diploid loci. An
individual having an allele assigned to the first-order group is
assigned to that group. Otherwise, the individual is assigned to the
next appropriate group. Two individuals sharing one allele might be
placed in either the same group or different groups, depending on the
mapping of alleles to groups in a particular iterate. For example,
consider how one might group two individuals, one with alleles {\em A1}
and {\em A2} at some locus, and another with alleles {\em A2} and {\em
A3}. Whether or not they are grouped together depends on the assignment
of alleles to groups, and can be done several different ways. The
algorithm enumerates each possible assignment of alleles to groups.

The extent of the search scales as $k^N$. In practice, the most
diverse locus was HLA-B, with 30 alleles when analyzed using two-digit
allele designations. For two groups, this gives $2^{30} \approx
10^{8}$ possible partitions. Serial iteration on an UltraSPARC-IIi
440MHz CPU (Sun Microsystems) requires roughly 36 hours for
completion. A parallel implementation requires no message passing, so
computing time scales inversely with an increasing number of CPUs, or
doubling available processors halves the time for iteration. With
many CPUs, the search space of $2^{30}$ partitions can be exhaustively
evaluated in an hour or less. Unfortunately, exhaustively evaluating
all three-way partitions is prohibitive, as $3^{30} \approx 2 \times
10^{14}$, over a million-fold increase in computational effort!
Supertype classification reduced the diversity of possible partitions
and enabled partitioning of the data into more than two groups.

The algorithm was implemented in C and will be distributed on request.

\subsection*{Class I \& II HLA Results}

The description length for the entire cohort as one group is $L_1=934$
bits; for the Caucasian subsample, it is $L_1=721$ bits. In general,
$L_1 < L_2$ at most loci (Table~1), so the MDL criterion does not
support partitioning alleles into groups that are predictive of high
or low RNA levels, except at HLA-B, where $L_2 < L_1$. In the
subsample, partitioning HLA-C or HLA-DQB1 alleles can also provide
preferred two-way splits, though not as well as HLA-B.  Further
partitioning was intractable because of great allelic diversity, as
previously mentioned. Partitions of HLA-B alleles provide the best
groupings among all loci. Because $L_2^* < L_1$, two groups,
partitioned by HLA-B alleles, provide a better description than one
(Fig.~1a and 1b).

What is the composition of the optimum groupings?  For the entire
cohort, the following alleles were associated with low viral RNA
levels: {\em B*13}, {\em B*27}, {\em B*38}, {\em B*45}, {\em B*49},
{\em B*57}, {\em B*58}, and {\em B*81}. The remaining alleles,
associated with greater viral RNA than the first group, are: {\em
B*07}, {\em B*08}, {\em B*14}, {\em B*15}, {\em B*18}, {\em B*35},
{\em B*37}, {\em B*39}, {\em B*40}, {\em B*41}, {\em B*42}, {\em
B*44}, {\em B*47}, {\em B*48}, {\em B*50}, {\em B*51}, {\em B*52},
{\em B*53}, {\em B*55}, {\em B*56}, {\em B*67}, and {\em B*82}. As
described earlier, having any alleles associated with the first group
is sufficient for an individual to be assigned to the group having
lower viral RNA.

How robust are these assignments of alleles to groups?  Four
alternative groupings provide description lengths within one bit of
the optimum. They do not dramatically rearrange the assigment of
individuals to groups, but do provide insight as to which alleles are
assigned to either group with less confidence. Among near-optimal
partitions, alleles {\em B*82} and {\em B*67} were assigned to groups
other than in the optimum partition.

In the Caucasian subsample, alleles {\em B*13}, {\em B*27}, {\em
B*40}, {\em B*45}, {\em B*48}, {\em B*49}, {\em B*57}, and {\em B*58}
are associated with lower viral RNA, and the remaining alleles, {\em
B*07}, {\em B*08}, {\em B*14}, {\em B*15}, {\em B*18}, {\em B*35},
{\em B*37}, {\em B*38}, {\em B*39}, {\em B*41}, {\em B*44}, {\em
B*47}, {\em B*50}, {\em B*51}, {\em B*52}, {\em B*53}, {\em B*55}, and
{\em B*56}, or lack of any alleles from the first group, are
associated with greater viral RNA levels. Two nearly optimal
partitions assigned alleles {\em B*47} and {\em B*48} to the second
group.  Fig.~1 illustrates the distributions of viral RNA levels from
this subsample, as one group (Fig.~1c) and as the best partition at
HLA-B (Fig.~1d).

To summarize the most robust inferences from the analyses of two-digit
allele designations, individuals having HLA-B alleles {\em B*13}, {\em
B*27}, {\em B*45}, {\em B*49}, {\em B*57}, or {\em B*58} were
associated with lower viral RNA levels than their counterparts lacking
these alleles.

Comparison of groupings obtained via the MDL approach with more
traditional means for statistical inference, a two-tailed, two-sample,
Welch modified t-test, which does not assume equal variances, and its
non-parametric variant, the Wilcoxon rank-sum test \cite{Venables},
was very favorable. In each case, the null hypothesis was that of no
difference between the group mean log-transformed viral RNA levels,
and the alternative hypothesis was that the means differ. Both tests
agreed in rejecting the null hypothesis in favor of the alternative
($P<10^{-10}$).

\subsection*{HLA Supertype Results}

Assigning the diploid, co-dominantly expressed HLA-A alleles to four
HLA-A supertypes \cite{supertypes}, {\em A1s}, {\em A2s}, {\em A3s},
and {\em A24s}, was possible for 399 individuals. The mapping of HLA-B
alleles to five supertypes, {\em B7s}, {\em B27s}, {\em B44s}, {\em
B58s}, and {\em B62s}, was made for 352 individuals. The resulting
decrease in allelic diversity enabled analysis for $k>2$.

Description lengths of the best $k$-way partitions of supertype
alleles for HLA-A supertypes are: $L_1=793$, $L_2=782$, $L_3=789$, and
$L_4=794$ bits. The best description length results from a two-way
split, though a three-way split also yields a shorter description
length than that obtained from one group. The best partition of HLA-A
supertypes assigned individuals having {\em A1s} alleles to the low
RNA group.

For HLA-B supertypes, $L_1=704$, $L_2=691$, $L_3=693$, and $L_4=697$
bits (Fig.~1e). The best model results when $k=2$. Overall,
individuals lacking {\em B58s} alleles averaged viral RNA levels
$3.6$-times greater than individuals having {\em B58s} supertype
alleles (Fig.~1f). Thus, individuals with {\em B58s} alleles have
significantly lower viral RNA levels than individuals without them.

Table~2 summarizes results of assigning HLA-B associations to high or
low viral-RNA categories as two-digit allele designations from both
the entire cohort and the Caucasian subsample, and as supertypes for
those individuals having two alleles that could be assigned to a
supertype. Alleles not found in a sample are indicated by a dash. The
{\em B*15} alleles are not shown because their high-resolution
genotype designations correspond to four different supertypes.

Overall, the most consistent associations with low viral RNA are among
the {\em B58s}, and with high viral RNA, the {\em
B7s}. Inconsistencies in assignment to a category occur for the {\em
B*13}, {\em B*27}, {\em B*45}, and {\em B*49} alleles, which are in
the low viral-RNA group when analyzed as such, but the high viral-RNA
group when assigned to supertypes.

When compared with alternative inferential techniques, the difference
between group viral RNA levels was highly significant. This and
agreement with alleles reported to be associated with variation in
viral RNA levels in previously published studies indicate that using
the description length as a test statistic can provide reliable
inferences.

\subsection*{MDL \& Statistical Inference}

The traditional statistical solution is to pose a question as follows:
suppose that the simpler model (e.g., one homogeneous population) were
actually true; call this the null hypothesis. How often would one, in
similar experiments, get data that look as different from that
expected under the null hypothesis as in the actual experiment?

This technique has limitations when the partition that represents the
alternative hypothesis is not given in advance. There are then many
potential alternative partitions and the appropriate distribution
under the null hypothesis for this ensemble of tests is very difficult
to estimate. Furthermore, for proper interpretation, the outcome
relies upon the truth of the initial assumption: that the data are
distributed as dictated by the null hypthothesis.

An alternative is to choose that model that represents the data most
efficiently. Here, efficiency is the amount of information, quantified
as bits, required to transmit electronically both the model and the
data as encoded by the model. This criterion may not seem intuitively
clear on first exposure. However, it follows naturally from a profound
relationship between probability and coding theory that was
discovered, explored, and elaborated by Solomonoff, Kolmogorov,
Chaitin, and Rissanen
\cite{Kolmogorov65,Chaitin66,Chaitin87,Rissanen86,Rissanen99}.

The idea is quite simple and elegant. It can be illustrated by analogy
to the problem of designing an optimal code for the efficient
transmission of natural-language messages. Consider the international
Morse code. Recall that Morse code assigns letters of the Roman
alphabet to codewords comprised of dots (``$\cdot$'') and dashes
(``$-$''). The codewords do not all have the same number of dots
and/or dashes; it is a variable-length code.

Efficient, compact encodings result from the design of a codebook such
that the shortest codewords are assigned to the most frequently
encoded letters and long codewords are assigned to rare letters. Thus,
{\em e} and {\em t} are encoded as ``$\cdot$'' and ``$-$'',
respectively, while {\em q} and {\em j} are encoded as ``$-$ $-$
$\cdot$ $-$'' and ``$\cdot$ $-$ $-$ $-$''. The theory of optimal
coding provides an exact relationship between frequency and code
length and thus, probability and description length.

The key departure of MDL from Morse-codelike schemes is that, while
Morse code would generally be good for sending messages over an
average of many texts, specific texts might be encoded even more
efficiently, by encoding not only letters, but letter combinations,
common words, or even phrases, perhaps as abbreviations or
acronyms. However, if one is to recode for particular texts, one must
first transmit the coding scheme. So perhaps one might use Morse code
to transmit the details of the new coding scheme and then transmit the
text itself with the new scheme. Whether this might yield greater
efficiency depends not only on how much compression is achieved in the
new encoding, but also on how much overhead is incurred in having to
transmit the coding scheme.

The analogy to scientific data analysis is clear. A statistical model
is an encoding scheme that encapsulates the regularities in the data
to yield a concise representation thereof. The best model effectively
compresses regularities in the data, but is not so elaborate that its
own description demands a great deal of information to be encoded.
The MDL principle provides a model-selection criterion that balances
the need for a model that is both appropriate and parsimonious, by
penalizing with equal weights the information required to specify the
model and the unexplained, or residual error.



Yet another contribution the MDL principle brings to statistical
modelling is that the penalty for multiple comparisons is less
restrictive than the penalty of compounded error rates incurred with
canonical inferential approaches. In order to maintain a desired
experiment-wide error rate, the standard adjustment is to make the
per-comparison error rate considerably more stringent. With current
technology, realistic sample sizes for such studies will generally be
less than a thousand and stringent significance levels will be
difficult to surpass. Unfortunately, fixing the false-positive error
rate does not address the false-negative probability, which may leave
researchers powerless to detect effects among many competing
hypotheses with limited samples.

\subsection*{Mechanisms}

Of HLA supertype alleles, individuals with {\em B58s} have lower viral
RNA levels than those who lack them, even among homozygotic
individuals. Naturally, this leads one to consider mechanisms that
underlie patterns found in the data. Elsewhere, we consider two
hypotheses to explain the observed associations between HLA alleles
and variation in viral RNA \cite{otherMS}.

There may be allele-specific variation in antigen-binding
specificity. Some alleles may have greater affinity than others for
HIV-specific peptide fragments due to the peptide-binding anchor
motifs they present. We were not able to identify any clear
association between the frequency of anchor motifs among HIV-1
proteins and viral RNA levels in the Chicago MACS \cite{otherMS},
though others have suggested that such a relationship might exist
\cite{Nelson}.

It may also the case that frequency-dependent selection has favored
rare alleles. Frequent alleles provide the evolving pathogen greater
opportunity to explore mutant phenotypes that may escape detection by
the host's immune response. By encountering rare alleles less
frequently, the virus has not had the same opportunity to explore
mutations that evade the host's defense response. This hypothesis is
corroborated by a significant association between viral RNA and HLA
allele frequency in the Chicago MACS sample \cite{otherMS}.

Because their predictions differ, these hypotheses could be tested
with data from another cohort, where a different viral subtype
predominates. That is, if other alleles were associated with low viral
RNA than those identified in this study, and an association between
rare alleles and low viral RNA levels were observed there, then the
second hypothesis would be more viable than the first. Alternatively,
if a clear association between antigen peptide-binding anchor motifs
and variation in viral RNA levels were found, the first hypothesis
would be more viable. Other mechanisms are also possible, and
hypotheses by which to evaluate them merit consideration.

\subsection*{Acknowledgments}

We thank Bob Funkhouser, Cristina Sollars, and Elizabeth Hayes for
sharing their expertise, and researchers of the Santa Fe Institute for
insight and inspiration. This research was financed by funds from the
Elizabeth Glazer Pediatric AIDS Foundation, the National Cancer
Institute, the National Institute of Allergy and Infectious Diseases,
National Institutes of Health, National Science Foundation award
\#0077503, and the U.S. Department of Energy.  We have no conflicting
interests.

\subsection*{Appendix}

In Gaussian Process modeling \cite{Williams97}, the population means
are treated as random variables and integrated out of the
likelihood. The model is then specified entirely by the structure of
the covariance matrix $\Sigma$, which specifies how each pair of
observations is correlated. The covariance is greater for two
observations from the same partition than for two observations from
different partitions. Any given partition is specified entirely by a
corresponding covariance structure.

{\bf Partitioning with Gaussian Models.}  Denote the $n$ observations
as the vector $Y$ and the covariance matrix with parameter vector
$\theta$ by $\Sigma(\theta)$. Let the number of components of $\theta$
(the number of free parameters in the covariance matrix) be $k$.  Then
the MDL for the given covariance structure is: 
$L = \frac{1}{2}\log |\Sigma(\hat{\theta})| + \frac{1}{2}
Y'\Sigma(\hat{\theta})^{-1}Y + \frac{k}{2}\log n + C$, where $C$ is
the information required to specify the partition or, equivalently,
the covariance structure, and $\hat{\theta}$ is the vector of
covariance parameters evaluated at maximum likelihood.

{\bf One Gaussian Population.}  The covariance matrix has a component
$\sigma ^2 _m$ for the covariance among observations, induced by their
sharing an unspecified mean, and an error component $\sigma ^2
_\varepsilon$: $\Sigma = \sigma ^2 _\varepsilon I + \sigma ^2 _m
\mathbf{11}'$, with $\mathbf{1}$ the column vector of all ones,
$\mathbf{11}'$ the matrix of all ones, and $I$ the identity
matrix. The inverse is:


\[\Sigma^{-1}= \frac{1}{\sigma ^2 _\varepsilon} \left( I - \frac{\sigma ^2 _m}{\sigma ^2 _\varepsilon + n \sigma ^2 _m} \mathbf{11}' \right), \]

and the log-determinant: $\log |\Sigma | = (n-1)\log \sigma ^2 _\varepsilon + \log (\sigma ^2 _\varepsilon + n \sigma ^2 _m)$.

This gives $L = \frac{1}{2} \left(n + (n-1)\log \sigma ^2 _\varepsilon + \log(\sigma ^2 _\varepsilon + n\sigma^2 _m ) + 2 \log n \right)$.

We find the maximum likelihood values of the parameters by minimizing
over the description lengths. There are two cases.\\

Case 1: $n^2 \overline{Y}^2 - Y'Y \ge 0$. Here we have 
$\hat{\sigma}^2 _\varepsilon =(n-1)^{-1}(Y'Y-n\overline{Y}^2)$ and
$\hat{\sigma}^2 _m =(n-1)^{-1}(n\overline{Y}^2 -\frac{1}{n}Y'Y)$, so 
$L=\frac{1}{2}(n +(n-1)\log\hat{\sigma}^2 _\varepsilon+ \log n \overline{Y}^2 + 2\log n)$.\\

Case 2: $n^2 \overline{Y}^2 - Y'Y < 0$. Here the common mean vanishes, giving
$\hat{\sigma} ^2 _\varepsilon = \frac{1}{n}Y'Y$, $\hat{\sigma}^2 _m = 0$, so 
$L = \frac{n}{2}(1+\log \hat{\sigma}^2 _\varepsilon +\frac{2}{n}\log n)$.\\
	 
{\bf Many Gaussian Populations.}
Two partitions give two populations. To analyze the HLA/HIV data,
we treated these populations as independent.  That is, we take the
covariance between observations in separate partitions to be zero, and
apply the fitting procedure outlined above separately to the two
populations. An alternative is to take non-zero covariance between the
two populations. This results in a more elaborate estimation
procedure, unlikely to yield large efficiency gains because the
two degrees of freedom (population means) are essentially mixed into
one, with residual error.

The procedure examines each admissible partition and computes the MDL
for that partition as the sum of individual description lengths over
the two independent populations. The best partition yields the lowest
description length over all partitions. This, plus the cost of
specifying the partition, is compared with the MDL from the
unpartitioned data. If the best partition provides a better
representation of the data than the unpartitioned set ($L _{k} < L
_{k-1}$), then the process is repeated in a recursive manner,
independently within each of the partitioned populations.

\newpage
\section*{Figure Legends}

Fig.~1. Description-length comparisons of viral RNA distributions as
one ($L_1$) or two ($L_2$) groups. Ordinate units are the expected
number of observations between two tick marks over the abscissa, or
one doubling of viral RNA. Impulses along the abscissa show individual
observations, with jitter added to enhance rendering of identical
values. (a) Observations ($n$) from the Chicago MACS cohort lumped
into one group, and (b) split into the best partition as two groups,
with individuals having alleles {\em B*13}, {\em B*27}, {\em B*38},
{\em B*45}, {\em B*49}, {\em B*57}, {\em B*58}, or {\em B*81} assigned
to the lower group ($n_1$), and remaining individuals assigned to the
group with greater viral RNA ($n_2$). (c) Observations from the
Caucasian subsample as one group, and (d) as the best split into two
groups, where having alleles {\em B*13}, {\em B*27}, {\em B*40}, {\em
B*45}, {\em B*48}, {\em B*49}, {\em B*57}, or {\em B*58} was the
criterion for being assigned to the low viral-RNA group. Observations
from individuals having two HLA-B supertype alleles, (e) in one group,
and (f) partitioned into two groups, contingent on the presence of
{\em B58s}.

\clearpage
\renewcommand{\baselinestretch}{2}

\begin{table}[tb]
\begin{center}
\caption{Optimum two-way partitions at each locus, with per-locus 
allelic diversity ($N$), description lengths without the information
cost to specify model parameters ($L_2 - C$), and minimum description 
lengths ($L_2$).
\label{tab:all-loci}}
\begin{tabular}{ccrcclcrccl}
\\
\hline
&\multicolumn{5}{c}{\sc{Entire Cohort}} & \multicolumn{5}{c}{\sc{Caucasian Subsample}}\\
& \multicolumn{5}{c}{$n=479$, $L_1=934$} & \multicolumn{5}{c}{$n=379$, $L_1=721$}\\
{\sc Locus}& & $N$ & $L_2-C$ && \multicolumn{1}{c}{$L_2$} & & $N$ & $L_2-C$ && \multicolumn{1}{c}{$L_2$}\\
\hline
\sc{Class I}\\
HLA-A && 19 & 916 && 935 && 18 & 703 && 721\\
HLA-B && 30 & 887 && 917* && 26 & 681 && 707*\\
HLA-C && 14 & 921 && 935 && 13 & 706 && 719\\
\sc{Class II}\\
DRB1  && 13 & 927 && 940 && 13 & 711 && 724\\
DQB1  &&  5 & 936 && 941 &&  5 & 715 && 720\\
DPB1  && 24 & 927 && 951 && 21 & 710 && 731\\
\hline
\end{tabular}
\end{center}
\end{table}
\newpage
\clearpage

\renewcommand{\baselinestretch}{1}

\begin{table}[t]
\begin{center}
\caption{HLA-B alleles associated with low ($\circ$) or high ($\bullet$) viral RNA levels.
\label{tab:hlab-summary}}
\begin{tabular}{cccc}
\\
\hline
             & {\sc Entire} & {\sc Caucasian} & {\sc Supertypes}\\
{\sc Allele} & {\sc Cohort} & {\sc Subsample} &    {\sc Only}\\
             &    $n=479$   &     $n=379$     &     $n=352$\\
\hline
\multicolumn{4}{l}{{\em B7s}}\\
{\em B*07}   & $\bullet$  & $\bullet$  &  $\bullet$\\
{\em B*35}   & $\bullet$  & $\bullet$  &  $\bullet$\\
{\em B*51}   & $\bullet$  & $\bullet$  &  $\bullet$\\
{\em B*53}   & $\bullet$  & $\bullet$  &  $\bullet$\\
{\em B*55}   & $\bullet$  & $\bullet$  &  $\bullet$\\
{\em B*56}   & $\bullet$  & $\bullet$  &  $\bullet$\\
{\em B*67}   & $\circ /\bullet$ & --   &  $\bullet$\\
\multicolumn{4}{l}{{\em B27s}}\\
{\em B*14}   & $\bullet$  & $\bullet$  & $\bullet$\\
{\em B*27}   & $\circ$    & $\circ$    & $\bullet$\\
{\em B*38}   & $\circ$    & $\bullet$  & $\bullet$\\
{\em B*39}   & $\bullet$  & $\bullet$  & $\bullet$\\
{\em B*48}   & $\circ /\bullet$  & $\circ /\bullet$ & $\bullet$\\
\multicolumn{4}{l}{{\em B44s}}\\
{\em B*18}   & $\bullet$  & $\bullet$  & $\bullet$\\
{\em B*37}   & $\bullet$  & $\bullet$  & $\bullet$\\
{\em B*40}   & $\bullet$  & $\circ$    & $\bullet$\\
{\em B*41}   & $\bullet$  & $\bullet$  & $\bullet$\\
{\em B*44}   & $\bullet$  & $\bullet$  & $\bullet$\\
{\em B*45}   & $\circ$    & $\circ$    & $\bullet$\\
{\em B*49}   & $\circ$    & $\circ$    & $\bullet$\\
{\em B*50}   & $\bullet$  & $\bullet$  & $\bullet$\\
\multicolumn{4}{l}{{\em B58s}}\\
{\em B*57}   & $\circ$    & $\circ$    & $\circ$\\
{\em B*58}   & $\circ$    & $\circ$    & $\circ$\\
\multicolumn{4}{l}{{\em B62s}}\\
{\em B*13}   & $\circ$    & $\circ$    & $\bullet$\\
{\em B*52}   & $\bullet$  & $\bullet$  & $\bullet$\\
\multicolumn{4}{l}{\sc Other}\\
{\em B*08}   & $\bullet$  & $\bullet$  & --\\
{\em B*15}   & $\bullet$  & $\bullet$  & --\\
{\em B*42}   & $\bullet$  & --         & --\\
{\em B*47}   & $\bullet$  & $\circ /\bullet$ & --\\
{\em B*81}   & $\circ$    & --         & --\\
{\em B*82}   & $\circ /\bullet$ & --   & --\\
\hline
\end{tabular}
\end{center}
\end{table}

\renewcommand{\baselinestretch}{1}

\clearpage

\thispagestyle{empty}
\setlength{\textwidth}{20cm}
\setlength{\headheight}{0cm}
\setlength{\headsep}{0cm}
\setlength{\topmargin}{0cm}
\setlength{\oddsidemargin}{0cm}
\setlength{\evensidemargin}{0cm}
\begin{figure}[p!]
\begin{center}
\epsfig{file=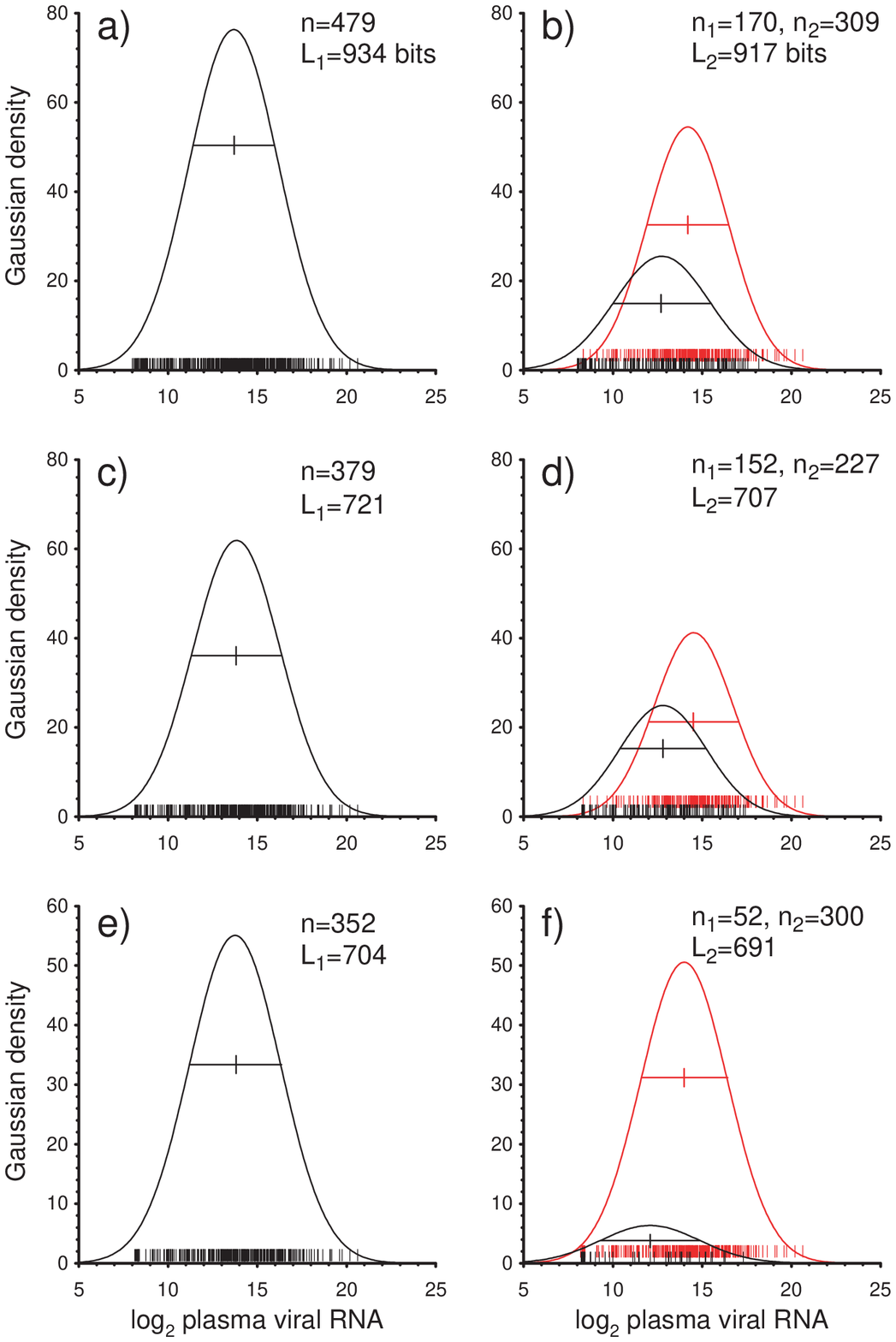,width=18cm}
\end{center}
\end{figure}
\clearpage

\end{document}